\documentstyle[12pt,epsfig]{article}
\newcommand{\be}{\begin{equation}}
\newcommand{\ee}{\end{equation}}
\newcommand{\al}{\mbox{$\alpha$}}

\newcommand{\bi}[1]{\bibitem{#1}}
\newcommand{\fr}[2]{\frac{#1}{#2}}

\newcommand{\R}{\mbox{$\fr{1-\gamma_5}{2}$}}
\newcommand{\GD}{\mbox{$\tilde{G}$}}

\newcommand{\Ima}{\mbox{Im}}

\begin{document}
\begin{titlepage} 
\rightline{UMN--TH--1827}
\rightline{UQAM-PHE-01-01}
\rightline{hep-ph/0105270}
\rightline{May 2001}  
\begin{center}

\vspace{0.5cm}

\Large
{\bf How natural is a small $\bar\theta$ in left-right SUSY models ? }
\normalsize

{\bf Cherif Hamzaoui$^{\bf a}$}\footnote{hamzaoui@mercure.phy.uqam.ca} and  {\bf 
Maxim 
Pospelov${\bf ^b}$}\footnote{pospelov@tpi1.hep.umn.edu}

\vskip0.3in  
\smallskip 
\medskip 
 \medskip 

{{\bf a} \it
D\'epartement de Physique, Universit\'e du Qu\'ebec \`a
Montr\'eal,\\ Case Postale 8888, Succ. Centre-Ville, Montr\'eal,
Qu\'ebec, Canada, H3C 3P8}

{{\bf b} \it Theoretical Physics Institute, School of Physics and Astronomy, 
University of Minnesota, Minneapolis, MN 55455, USA} 
\smallskip 
\end{center} 
\vskip1.0in 
 
\noindent{\large\bf Abstract} 
\smallskip\newline
In the world without an axion 
the smallness of $\bar\theta$ may be achieved due to a spontaneously 
broken discrete left-right symmetry. 
We analyze the radiatively induced $\bar\theta$ in the context of 
generic left-right symmetric SUSY models  
without assuming flavor degeneracy 
in the squark sector. Left-right symmetry allows to keep 
$\bar\theta$ within its present bound only if the inter-generational 
mass splitting in the squark sector at the scale of the left-right symmetry 
breaking is smaller than 0.5\%. We also consider the naturalness of $m_u=0$
solution to the strong CP problem in the context of horizontal flavor
symmetries. A strong bound on the combination of the horizontal charges
in the Up quark sector is found in this case.

\end{titlepage} 

\section{Introduction} 

The strong CP problem remains an important and open issue in particle 
physics. The QCD Lagrangian has a fundamental
parameter $\theta$ which labels 
different super-selection sectors. Its effect can be accounted by
an additional term in the QCD Lagrangian,
\be
{\cal L}= \theta\fr{g^2_3}{16\pi^2} G^a_{\mu\nu}\GD^a_{\mu\nu},
\ee
which  violates P and CP symmetries \cite{theta}.
 
In a full theory, the effective low energy value for the 
theta term is modified by possible complex phases in the quark mass matrices:
\be
\bar{\theta}=\theta+{\rm arg}({\rm det} M_uM_d)+...
\label{mumd}
\ee
The ellipsis stands for other possible contributions from yet unknown
fermions, charged with respect to $SU(3)$ color gauge group (for example, 
gluino). 

Current experimental limits on the electric dipole moment (EDM) of the 
neutron put severe constraints on the allowed size of the 
$\bar{\theta}$-angle. Among different ways of calculating the EDM
of the neutron, induced by $\bar\theta$, 
the most reliable ones use chiral 
perturbation theory \cite{CDVW} or QCD sum rules \cite{PR1,PR2,Henley}.  
Here we use the numerical result of ref. \cite{PR1} which
gives the following prediction for $d_n(\bar\theta)$:
\be
d_n\simeq 1.2\times 10^{-16}\bar{\theta}\,e\cdot cm.
\ee
Together with the current experimental limits on the neutron EDM  \cite{nEDM}
it implies a stringent bound on the theta term, 
$\bar{\theta}<6\times 10^{-10}$. Similar bounds are provided by the 
limits on the electric dipole moment of $^{199}Hg$ atom \cite{Mike,HH,KL}. 

The puzzling smallness of  $\bar{\theta}$ in comparison 
with a natural expectation of $\bar\theta\sim 1$ is usually referred 
to as the strong CP problem. There are several theoretical possibilities
of removing $\bar\theta$ from the theory, none of 
which are free from their own intrinsic problems.  
The most popular solution to the strong CP problem uses
the dynamical relaxation of $\bar{\theta}$ through 
the axion mechanism \cite{PQ}. Perhaps, it is the most 
elegant and universal way of removing the theta term. 
However, the negative results of all experimental
searches of axions and very restrictive astrophysical and cosmological 
considerations which place the axion coupling constant into a relatively
narrow range, $10^{10}-10^{12}$ GeV, suggest to look for other alternative
solutions.

In principle, one can speculate on the vanishing of the Yukawa 
coupling for the up quark, hoping that the hadronic phenomenology
would still allow for $m_u=0$ \cite{KM,CKN}. 
The vanishing of this coupling may be a consequence of the horizontal flavor
symmetries, supposedly responsible for the hierarchy of
masses and mixing angles in the fermion sector \cite{NS,BNS,MNR}. We will 
comment on this possibility in supersymmetric models 
and show that $m_u$ is highly susceptible to the 
supersymmetric threshold corrections.  
$m_u=0$ is unnatural unless there exists large differences in
horizontal charges in the Up sector. 

The main goal of this paper is to consider an important class of 
solutions where $\bar\theta$  is small {\em by construction}. This 
can occur if parity or CP symmetry are exact symmetries
of the full theory. Here we assume that these symmetries are 
spontaneously broken at some energy scale above the electroweak scale. 
In the absence of the axion mechanism, 
the radiative corrections to $\bar \theta$ which are induced below this scale,
will be the main source of the EDMs in the hadronic sector. Therefore,
these corrections have to be within the experimental bounds. 
This provides severe restrictions on the amount of 
CP-violation that one can have in this class of theories \cite{DP}.

The models with spontaneously broken CP, constructed 
to solve the strong CP problem \cite{CP}, 
normally fail to keep $\bar \theta$ within 
the experimental bound  after the radiative corrections are taken 
into account. In the supersymmetric 
framework this problem  was emphasized in Ref. \cite{DKL}, 
where it was shown that the non-universality in the squark sector 
generates $\theta_{rad}$ 
considerably larger  than the experimental bound. 
In order to keep the radiative corrections to $\bar \theta$ small, one
has to suppress all CP-violating phases in the theory, including the phase in 
the Kobayashi-Maskawa matrix. In general, it is hard to achieve and 
these  are of the superweak type models. 
Recent confirmation of the non-zero result on $\epsilon'/\epsilon$ \cite{eps}
casts strong doubts on the viability of the 
superweak framework and disfavors most of the models with 
small $\bar\theta$ due to spontaneous breaking of CP. 
There is, however, a recent model-building proposal
which combines the low energy SUSY breaking and strong 
complex renormalization of the quark wave functions
at high scale with unbroken SUSY in order to get zero $\bar\theta$ {\em and} 
generate sufficiently large Kobayashi-Maskawa phase \cite{CP1}. 

The idea of the spontaneous breaking of parity, initially introduced in
the framework of Pati-Salam unification \cite{PS}, was thoroughly
studied in the case of $SU(3)\times SU(2)_L\times SU(2)_R\times U(1)$ gauge 
group \cite{LR}. Among different appealing features of these models 
is the possibility to have $\theta=0$
at the tree level as a consequence of exact left-right symmetry under which
$\theta \rightarrow -\theta$ \cite{LRCP}. Several years ago, this idea 
was discussed
again in the context of the supersymmetric left-right models \cite{K,MR}.
The vanishing of $\bar\theta$ at the tree level does not necessarily lead 
to a solution to the strong CP problem as loop corrections below the scale of 
left-right breaking can generate $\theta_{rad}$ above the required bound.
Careful analysis of the radiative corrections to the theta term 
performed in Ref. \cite{Posp} showed that $\theta_{rad}$ can be kept 
within the experimental limits, {\em assuming} 
the initial universality in the squark sector at the scale of the SUSY
breaking. However, it is not generally expected that the universality 
in the squark sector is exact, and therefore it is not clear how large the 
radiative corrections to $\bar\theta$ in a generic left-right SUSY model 
can be. One can hope that a moderate splitting among the squark 
masses would allow to keep $\bar\theta$ 
within the experimental bound. Examples of  
$\sim 1$\% splitting in the squark mass sector  
could be attained in some variants of the free fermionic 
superstring models \cite{FP,FD}. 

In this paper we explore the necessary 
conditions for naturally small $\bar\theta$ in a 
generic supersymmetric theory which has the following features. 
Below some scale $\Lambda$, the field content of the theory is that of the 
Minimal Supersymmetric Standard Model with the SM gauge group. Above 
$\Lambda$, the theory has the  
built-in discrete left-right symmetry and more complicated gauge structure 
compatible with it. It is important that the vanishing of the $\bar \theta$ 
parameter at the  tree level in these models relies 
on the existence of the discrete left-right symmetry rather than on the 
particular choice of the gauge group. We assume that the 
Higgs structure below the scale of the left-right breaking is minimal
in the spirit of ref. \cite{zurab}.
We analyze radiative corrections to the theta term 
without assuming squark degeneracy and 
find the allowed degree of the non-universality in 
the soft-breaking sector consistent with the bounds on theta. 
It turns out that EDMs require $0.5\%$ degeneracy in the squark sector 
at the scale $\Lambda$ as well as a strong alignment of squark masses 
and Yukawa couplings in the Down quark sector.

\section{Theta term and left-right symmetry in SUSY}

Previous works on the 
the theta problem in  $SU(3)\times SU(2)_L\times SU(2)_R\times U(1)$ 
left-right 
symmetric theories \cite{K,MR,Posp} used
very specific ansatz of proportionality
and degeneracy in the squark sector. 
By these conditions we understand the following 
requirements imposed on the soft-breaking sector at the scale of the
SUSY breaking:
\begin{eqnarray}  
{\bf M}_Q^2= m_Q^2{\bf 1};\;\;{\bf M}_D^2= m_D^2{\bf 1}; \;\;{\bf M}_U^2=
m_U^2{\bf 1}\;\;\;\mbox{"degeneracy"} 
\label{eq:deg}\\
{\bf A}_u= A_u{\bf Y}_u;\;\;{\bf A}_d= A_d {\bf Y}_d\;\;\;%
\mbox{"proportionality"}.   \label{eq:prop}
\end{eqnarray}

It is known, however, that in the models which use a spontaneously broken CP
symmetry to solve strong CP problem, the departure from the exact universality
should be at a very tiny level, $10^{-6}$ or so \cite{DKL}.
To determine allowed squark mass splittings in LR SUSY models, 
we relax the conditions of the universality and 
proportionality, while keeping explicit 
left-right symmetry at the scale $\Lambda$. This symmetry requires the 
hermiticity of the Yukawa matrices ${\bf Y}$ and trilinear scalar 
coupling matrices ${\bf A}$, as well as the identity of the left and right
handed squark mass matrices,
\begin{eqnarray}\nonumber
{\bf Y}_u={\bf Y}_u^\dagger,\;\;\;\;{\bf Y}_d={\bf Y}_d^\dagger,\;\;\;\;
{\bf A}_u={\bf A}_u^\dagger,\;\;\;\;{\bf A}_d={\bf A}_d^\dagger\newline\\
{\bf M}_Q^2={\bf M}_D^2={\bf M}_U^2={\bf M}^2\equiv m^2{\bf 1}+{\bf S}.
\label{eq:lr}
\end{eqnarray}
Matrix ${\bf S}$ parametrizes the departure from the degeneracy condition,
and we choose ${\bf S}$ in such a 
way that Tr${\bf S}=0$.  
Another important set of conditions is the absence of the explicit 
CP-violating phases in the gaugino masses and $B$-parameter in the 
Higgs potential,
\be
B=B^*,\;\;\;m_{\lambda_i}=m_{\lambda_i}^*,\;\;\; \mu=\mu^*.
\label{eq:noph}
\ee 
As noted in \cite{K,MR}, the reality of the $SU(2)$ gaugino mass
does not follow from the left-right symmetry. We consider it as a consequence
of a higher unification scheme which makes all gaugino phases equal.

Conditions (\ref{eq:lr}) and (\ref{eq:noph}) ensure the absence of 
the theta term 
at the tree level. At the loop level one has to consider radiative corrections
to the quark and gluino mass matrices. These corrections 
must be proportional to CP-violating phases present in the theory. 
When squarks are degenerate, this source is just the complexity of the 
Yukawa matrices, which is the Kobayashi-Maskawa (KM) phase in this case.
The latter provides a {\em minimal} content of 
CP-violation. If the contribution to $\bar{\theta}$ from KM phase 
happens to be 
large, this means that one cannot obtain a viable solution to the strong CP 
problem without fine tuning. This question was studied in the framework of 
pure SM \cite{EG,Kh}, where radiative corrections to $\bar{\theta}$ arise 
first in the order $\al_sG_F^2m_c^2m_s^2$ times the CP-odd KM invariant 
$I_{KM}$ \cite{Kh}, and in the MSSM with the KM mechanism of CP-violation 
\cite{DGH} where the result is also found to be much smaller than 
$10^{-9}$. The main reason behind the suppression of $\theta_{rad}$
is the smallness of the Yukawa couplings and mixing angles
contained in the so--called Jarlskog factor, 
\be
J_{SM}=I_{KM}(\lambda_t^2-
\lambda_c^2)(\lambda_c^2-\lambda_u^2)(\lambda_u^2-\lambda_t^2)
(\lambda_b^2-\lambda_s^2)(\lambda_s^2-\lambda_d^2)
(\lambda_d^2-\lambda_b^2)
\label{jarl}
\ee 
Here $I_{KM}$ is the imaginary part of the invariant quartic combination
of the KM mixing elements, $I_{KM}=\Ima(V^*_{td}V_{tb}V^*_{cb}V_{cd})$.

When the squark degeneracy is abandoned, an analog of 
Jarlskog--type factor may come from the squark sector. 
If violation of degeneracy is ``moderate'', 
i.e. proportional to the small
factor $\epsilon_{ij} =(s_i-s_j)/m^2$, there is a chance 
that the experimental bound on 
$\bar\theta$ can be satisfied without fine tuning. 

In the case of MSSM  the leading contributions to 
$\bar\theta$ come from corrections to
the quark mass matrix, gluino mass term and 
the radiative correction to the 
Higgs potential, Fig. 1. 

\begin{figure}[thb]
 \begin{center}
\epsfig{file=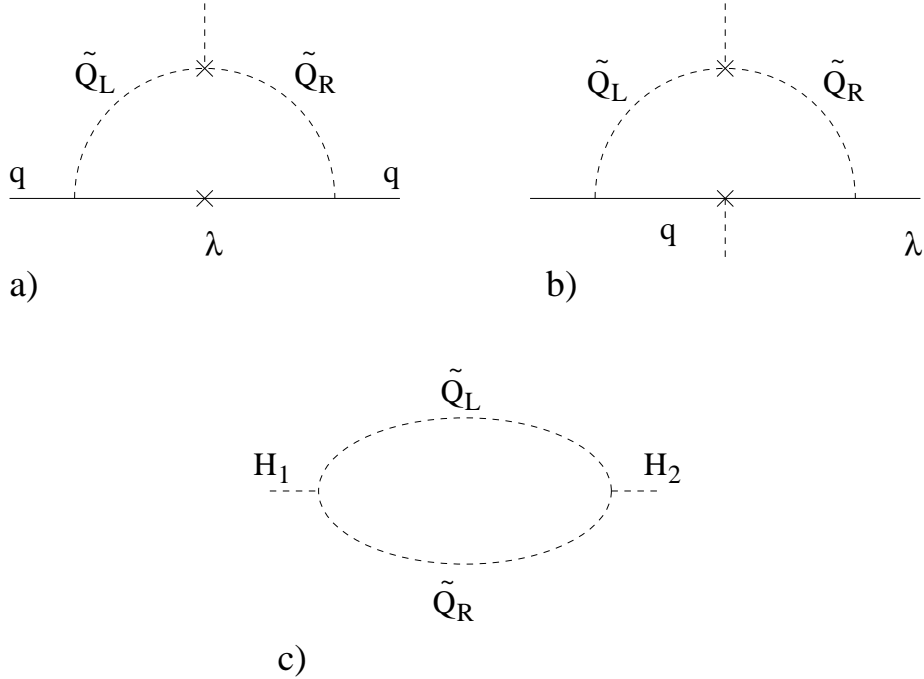,height=3.5in} 
\caption{ SUSY threshold corrections to {\em a)} quark Yukawa couplings,
{\em b)} gluino mass, {\em c)} $m_{12}^2$ soft breaking parameter which 
can complexify $v_1v_2$ }   \ \end{center}
\vspace{-0.3in}\end{figure}

As it will become clear shortly, the corrections to the quark masses are 
by far more important than other types of contributions. 
In order to solve strong CP problem, the corrections to 
Yukawa matrices which can complexify their determinants should be small 
and thus we expand eq. (\ref{mumd}) to obtain
\be
\bar\theta={\rm Im}\left[{\rm Tr}({\bf Y}_u^{-1}\Delta{\bf Y}_u)+
{\rm Tr}({\bf Y}_d^{-1}\Delta{\bf Y}_d)\right ].
\ee

Below the scale $\Lambda$, the general form of the squark mass matrices
becomes very complicated. The renormalization group equations are no
longer left-right symmetric and induce the splitting between left- and 
right-handed squarks; the Yukawa matrices enter into the running of the 
squark masses, etc. To analyze SUSY threshold corrections to $\bar\theta$ 
we choose the following strategy. We treat separately the cprrections
to $\bar\theta$ induced 
by the resulting left-right asymmetry in the squark masses and those induced
by  
the intergenerational splitting which is 
induced by Yukawa interactions. In both cases
we derive constraints on ${\bf S}$-matrix and require  ${\bf S}$ to 
satsify all of them, thus neglecting possible cancellations among 
different mechanisms of inducing $\theta$. Since we are interested in
{\em natural} solution to the strong CP problem, possible cancellations 
between different mechanisms should be rightfully ignored as they represent
a fine tuning which this solution wants to avoid. 
In addition to that, we treat  $\mu$ and ${\bf A}$-proportional 
corrections separately.

The gauge evolution of the squark masses 
from the scale $\Lambda$ to the electroweak scale
induce the left and right-handed squark mass splitting, 
$m_L^2-m_R^2 \simeq m_2^2 (3\al_2/2\pi) \ln({\Lambda^2/m^2)}$.  
In our final result  we take this splitting to be $O(1)$. 
The value of $\theta$ induced by this effect can be estimated 
by expanding the diagram, given by Fig. 1a, in ${\bf S}$ matrix 
and keeping only the lowest possible terms in ${\bf S}$. This procedure is 
justified as long as mass splittings in the squark sector are smaller 
than characteristic momenta in the loop. Thus we have
\begin{eqnarray}
\label{kl0}
{\rm Im}{\rm Tr}({\bf Y}_d^{-1}\Delta{\bf Y}_d)
=i\fr{8g_3^2}{3}
\int\fr{d^4p}{(2\pi)^4}\fr{\mu m_\lambda\tan\beta}{p^2-m_\lambda^2} 
\sum_{k,l}\fr{{\rm Im}{\rm Tr}({\bf Y}_d^{-1}{\bf S}^k{\bf Y}_d{\bf S}^l)}
{(p^2-m_L^2)^{k+1}(p^2-m_R^2)^{l+1}}\\
{\rm Im}{\rm Tr}({\bf Y}_u^{-1}\Delta{\bf Y}_u)
=i\fr{8g_3^2}{3}
\int\fr{d^4p}{(2\pi)^4}\fr{\mu m_\lambda\cot\beta}{p^2-m_\lambda^2} 
\sum_{k,l}\fr{{\rm Im}{\rm Tr}({\bf Y}_u^{-1}{\bf S}^k{\bf Y}_u{\bf S}^l)}
{(p^2-m_L^2)^{k+1}(p^2-m_R^2)^{l+1}}
\label{kl}
\end{eqnarray}

It is easy to see that the lowest possible order in which the CP violation
does not vanish is $k+l=3$. It is convenient to choose 
the basis in which squark masses are diagonal:
\be
{\bf S} = diag(s_1,s_2,s_3),\;\; 
{\bf Y}_u = U {\bf Y}_u^{diag} U^\dagger,\;\; 
{\bf Y}_d = V {\bf Y}_d^{diag} V^\dagger.
\ee
The product of $U$ and $V$ matrices gives the Kobayashi-Maskawa mixing matrix,
$K=U^\dagger V$. Then the two lowest order structures, naturally arising from
eqs. (\ref{kl0}-\ref{kl}), are
\begin{eqnarray}
J_u=\Ima (U_{21}^*U_{23}U_{33}^*U_{31})(s_1-s_2)(s_2-s_3)(s_3-s_1)
\fr{(m_u-m_c)(m_c-m_t)(m_t-m_u)}{m_um_cm_t}\nonumber\\
J_d=\Ima (V_{21}^*V_{23}V_{33}^*V_{31})(s_1-s_2)(s_2-s_3)(s_3-s_1)
\fr{(m_d-m_s)(m_s-m_b)(m_b-m_d)}{m_dm_sm_b}
\label{jujd}
\end{eqnarray}
To estimate the size of the integral we take $m_L\simeq m_R\simeq m_\lambda$
and arrive at the following expression for $\bar \theta$
$$
\bar\theta\simeq \fr{2\alpha_s}{90\pi}~\fr{(m_L^2-m_R^2)\mu}{m^3}
~\fr{J_u\cot\beta+J_d\tan\beta}{m^6}\simeq \nonumber
$$
\be
8.5\cdot 10^{-4}~\fr{(m_L^2-m_R^2)\mu}{m^3}~
\epsilon_{12}\epsilon_{23}\epsilon_{31}~\left(\fr{m_t}{m_u}
\Ima (U_{21}^*U_{23}U_{33}^*U_{31})\cot\beta+ \fr{m_b}{m_d}
\Ima (V_{21}^*V_{23}V_{33}^*V_{31})\tan\beta\right).
\ee
Here, only the leading terms enhanced by the ratios $m_t/m_u$ or $m_b/m_d$ 
are retained in eqs. (\ref{jujd}). We note in passing that 
the answer in this form does not allow to take the limit 
$m_{u(d)}\rightarrow 0$ because it relies on $\Delta m_u\sim 
\Ima(U...U)m_t \ll m_u$. 
The numerical bound on theta is satisfied as long as 
\begin{eqnarray}\label{limit0}
\fr{(m_L^2-m_R^2)\mu\tan\beta}{m^3}~\Ima (V_{21}^*V_{23}V_{33}^*V_{31})~
\epsilon_{12}\epsilon_{23}\epsilon_{31} < 10^{-9}\\
\fr{(m_L^2-m_R^2)\mu\cot\beta}{m^3}~\Ima (U_{21}^*U_{23}U_{33}^*U_{31})~
\epsilon_{12}\epsilon_{23}\epsilon_{31} < 10^{-11}
\label{limits}
\end{eqnarray}
Equations (\ref{limit0}-\ref{limits}) suggest that 
the squarks have to be degenerate at 
1\% level in the basis in which mixing angles of $U$ and $V$ are on the 
order of CKM mixing angles, $ U_{ij} \sim V_{ij} \sim K_{ij} $. Indeed, with 
$\Ima (V_{21}^*V_{23}V_{33}^*V_{31})\sim \Ima (V_{21}^*V_{23}V_{33}^*V_{31})
\sim I_{KM} \sim 10^{-5}$ and $\epsilon_{ij} \sim 0.01$, both constraints 
(\ref{limit0}) and (\ref{limits})
can be satisfied. The most relaxed constraints on $\epsilon_{ij}$, 
$\epsilon_{ij}< 0.05$ are for the case of $U= {\bf 1}$ and $V=K$, 
which corresponds to the squark matrix, ${\bf Y}_u$ being 
diagonal in the same basis. 

The inclusion of the squark mass renormalization group running, 
generated by the Yukawa interaction,
introduce additional ``dangerous'' corrections to $\bar\theta$. 
The change of the mass matrices at one loop level is given by
the following set of expressions (See, e.g. ref. \cite{HPR}), linearized 
in the renormalization group coefficients:
\begin{eqnarray}
M^2_{uLL}= m^2{\bf 1}+{\bf S} +{\bf M}^\dagger_u{\bf M}_u
+c_1{\bf Y}^\dagger_u(m^2{\bf 1}+{\bf S}){\bf Y}_u+
c_2{\bf Y}^\dagger_d(m^2{\bf 1}+{\bf S}){\bf Y}_d\\
M^2_{uRR}= m^2{\bf 1}+{\bf S} +{\bf M}_u{\bf M}_u^\dagger
+c_3{\bf Y}_u(m^2{\bf 1}+{\bf S}){\bf Y}_u^\dagger
\nonumber
\end{eqnarray}
For the down squark matrices, $u$ and $d$ indices should be interchanged. 
The renormalization group coefficients $c_i\sim
(16\pi^2)^{-1} \ln(\lambda^2/m^2)$ can be found in refs. \cite{MV,HPR}. 
 Their particular forms 
are not important for our purposes as we take them to be 
$O(1)$.
 Using these matrices we calculate the theta term, again expanding the 
 propagators in the Yukawa couplings. In principle, 
 for the top quark one should retain 
all orders in this expansion and the correct way of doing this was 
given in ref. \cite{HPR}. However, for the present discussion 
it is sufficient to keep only the first-order term. 
The most important contributions to the $\bar\theta$ parameter are coming from
the expansion of the right-handed squark line in ${\bf S}$ and 
left-handed squark line in ${\bf Y}_{u(d)}$:
\begin{eqnarray}
{\rm Im}{\rm Tr}({\bf Y}_d^{-1}\Delta{\bf Y}_d)
=i\fr{8g_3^2}{3}
\int\fr{d^4p}{(2\pi)^4}~\fr{m^2\mu m_\lambda\tan\beta}{p^2-m_\lambda^2} ~
\fr{c_2\Ima{\rm Tr}({\bf Y}_d^{-1}{\bf Y}_u^2{\bf Y}_d{\bf S})}
{(p^2-m^2)^4}\\
{\rm Im}{\rm Tr}({\bf Y}_u^{-1}\Delta{\bf Y}_u)
=i\fr{8g_3^2}{3}
\int\fr{d^4p}{(2\pi)^4}~\fr{m^2\mu m_\lambda\cot\beta}{p^2-m_\lambda^2}~ 
\fr{c_2\Ima{\rm Tr}({\bf Y}_u^{-1}{\bf Y}_d^2{\bf Y}_u{\bf S})}
{(p^2-m^2)^4}
\label{eq:yys}
\end{eqnarray}

Taking advantage of large mass ratios in the quark sector, we reduce these
expressions to a simpler form given by the combination of $K$, $V$ and
$U$ matrix elements:
\begin{eqnarray}
\label{strong}
\fr{\alpha_s}{18\pi}\fr{\mu\tan\beta}{m}c_2y_t^2\fr{m_b}{m_d}
\Ima\sum_{ij}K_{id} K^*_{td} K_{tb} K_{jb}^* 
(U\fr{S^{diag}}{m^2}U^\dagger)_{ji}\\
\fr{\alpha_s}{18\pi}\fr{\mu\cot\beta}{m}c_2y_b^2\fr{m_t}{m_u}
\Ima\sum_{ij}K^*_{ui} K_{ub} K^*_{tb} K_{tj} 
(V\fr{S^{diag}}{m^2}V^\dagger)_{ji}\label{strong1}
\end{eqnarray}

In a general case, allowed by previous constraints (\ref{limits}), 
$|U_{ij}|;\,|V_{ij}|\sim |K_{ij}|$ and $\epsilon_{ij}\sim 0.01$, the new 
contributions (\ref{strong}-\ref{strong1}) 
to $\bar\theta$ would violate the experimental bound. 
To satisfy it, we would have to impose strong restrictions 
on the splitting in the squark sector, $\epsilon_{ij}< 10^{-4}$. 
There are, however, two important exceptions from this constraint, 
which we should treat separately.

{\em Case 1:} this is when the squarks and Yukawa couplings of down quarks are 
diagonal in the same basis. This corresponds to $V= {\bf 1}$ and 
$U^\dagger=K$ and leads to the vanishing of the expression in 
eq. (\ref{strong}). The theta term is given by 
\be
\bar\theta = \fr{\alpha_s}{18\pi}\fr{\mu\cot\beta}{m}c_2y_b^2I_{KM}
\epsilon_{12}\simeq 10^{-7}\epsilon_{12}
\ee
As we can see, the value of $\epsilon_{12} = 0.005$ is consistent with 
the theta-constraint if the value of $\tan^\beta\sim O(1)$. 

{\em Case 2:} this is when the deviation from universality is expressed as 
the function of traceless bilinear combinations of Yukawa matrices.
\be
{\bf S} = am^2 ({\bf Y}_u^2- \fr{1}{3}{\rm Tr}\left({\bf Y}_u^2)\right) + b m^2
\left({\bf Y}_d^2- \fr{1}{3}{\rm Tr}({\bf Y}_d^2)\right) 
\label{peculiar}
\ee
This is the generalization of the model discussed in ref. \cite{Posp}.
Indeed, this form of the mass matrix at the scale $\Lambda$ can be 
viewed as the result of the squark universality at the Plank scale,
modified by the renormalization group flow {\em above} the scale $\Lambda$. 
Due to a higher left-right symmetric group, 
more Higgses are present and this explains the 
appearance of both $ {\bf Y}_d$  and $ {\bf Y}_u$ in Eq. (\ref{peculiar}).
In this case the resulting value of $\bar\theta$ was estimatyed in ref. 
\cite{Posp},
\be
\bar{\theta}= I_{KM}\fr{\al_s}{90\pi}
\fr{(a+c_2)ac_2\mu\tan\beta}{m}y_t^4y_c^2\fr{m_b}{m_d} \sim 10^{-10}
\tan\beta
\label{eq:anal}
\ee
This value is within a desirable bound if $\tan\beta$ is not too large. 

Next we include the corrections coming from the violation of
proportionality in ${\bf A}$-matrices. The hermiticity of ${\bf A}_{u(d)}$ at 
the scale $\Lambda$ is violated at lower scales so that 
\be
{\bf A}_u\rightarrow (1+c_4{\bf Y}_u^2 +
c_5{\bf Y}_d^2){\bf A}_u(1+c_6{\bf Y}_u^2),
\ee
with the similar transformation for ${\bf A}_d$. 
Insertion of these structures 
into the squark line leads to the following potentially dangerous 
corrections to the Yukawa matrices:
\begin{eqnarray}\label{ad}
{\rm Im}{\rm Tr}({\bf Y}_d^{-1}\Delta{\bf Y}_d)
=i\fr{8g_3^2}{3}
\int\fr{d^4p}{(2\pi)^4}~\fr{m_\lambda}{p^2-m_\lambda^2} ~
\fr{c_5\Ima{\rm Tr}({\bf Y}_d^{-1}{\bf Y}_u^2{\bf A}_d)}
{(p^2-m^2)^3}\\
{\rm Im}{\rm Tr}({\bf Y}_u^{-1}\Delta{\bf Y}_u)
=i\fr{8g_3^2}{3}
\int\fr{d^4p}{(2\pi)^4}~\fr{m_\lambda}{p^2-m_\lambda^2}~ 
\fr{c_5\Ima{\rm Tr}({\bf Y}_u^{-1}{\bf Y}_d^2{\bf A}_u)}
{(p^2-m^2)^3}
\label{au}
\end{eqnarray}
Let us parametrize ${\bf A}$-matrices in the basis where the same type 
(up or down) Yukawa matrices are diagonal:
\be
{\bf Y}_d =  {\bf Y}_d^{diag},\;\; 
{\bf A}_d = V_A {\bf A}_d^{diag} V_A^\dagger; \;\; \;\; 
{\bf Y}_u =  {\bf Y}_u^{diag},\;\; 
{\bf A}_u = U_A {\bf A}_u^{diag} U_A^\dagger
\ee
 Then the equations (\ref{ad}-\ref{au}) 
provide severe restrictions either on the allowed form of the
 $V_A$ and $U_A$ matrices or on the magnitude of the eigenvalues of 
$A$ matrices at the scale 
$\Lambda$. In particular, we find that for $  A_b^{diag} \sim my_b$, 
the allowed values of element 13 of $V_A$ matrix should be of the order
$O(10^{-7})$ or smaller. This suggests that the allowed departure 
from proportionality may occur only at the level of eigenvalues,
i.e. $A_b/y_b\neq A_s/y_s\neq A_d/y_d$, while $V_A=U_A=1$ must be preserved. 
Another, more radical assumption, would be to take the matrices ${\bf A}_u$ and
${\bf A}_d$ at the scale $\Lambda$ to be arbitrary but all entries 
suppressed to the level of $10^{-7}$.

\section{Strong CP, $m_u=0$ and horizontal symmetries}

  Horizontal symmetries have a potential to explain 
the hierarchical patterns among the 
quark masses and mixing angles \cite{FN}. Recently there 
have been some activities in supersymmetric models supplemented 
by horizontal symmetries.
These symmetries might be behind an approximate 
``alignment'' between squark and quark mass matrices \cite{NS}.

The basic idea in this approach is to relate the smallness of some
entries in the Yukawa matrices with certain powers of the order parameter
$\lambda=\langle s \rangle /M\ll 1$, 
characterizing the breaking of the horizontal
symmetry. In other words, below the scale $M$, the superpotential is the 
sum of different operators, classified by the dimension of $\lambda$, 
\be
W = \sum_{ij}Q_i U_j H_2c_i
\left(\fr{\langle s \rangle }{M}\right)^{p_{ij}}+...
\ee
Coefficients $c_i$ are of the order 1. 
Let us assume that the $H_2$ field is not charged with respect to 
the horizontal group and that $s$-field carries unit negative charge,
$X_s=-1$. Then the selection rule for $p_{ij}$ can be formulated
in the following form:
\be
p_{ij}= \left\{\begin{array}{c}
X_{Q_i}+X_{U_J}; \;\;\;\mbox{for}\; X_{Q_i}+X_{U_j}\geq 0\\
 0\;\;\;\; \mbox{for}\;            X_{Q_i}+X_{U_j}< 0.                   
                         \end{array}\right.
\label{selection}
\ee

Thus the holomorphic properties of the superpotential, i.e. the absence of 
terms like $QUH_2s^*$, can be used to decouple 
right-handed $u$-quark from the rest of the MSSM chiral superfields. 
In other words, an accidental $det Y_u=0$ can be a natural 
consequence of horizontal symmetries in 
the SUSY framework \cite{BNS}, thus 
reviving an $m_u=0$ solution to the strong CP problem. 
 
Does this solution withstand radiative corrections to $m_u$ 
which may arise due to the same SUSY threshold correction, Fig. 1a?
To answer this question we have to check whether it is natural to have
radiatively induced $m_u({\rm 1~GeV})<10^{-9}\cdot 5 MeV$. 
As it was pointed out in \cite{NS}, the selection rule for 
the squark mass matrix elements, $M_{ij}^2 \sim m^2 
(\langle s\rangle/M)^{q_{ij}}$ is quite different from 
(\ref{selection}),
\be
q_{ij}=|X_{Q_i}-X_{Q_j}|,\;\; |X_{U_i}-X_{U_j}|,\;\;  
|X_{D_i}-X_{D_j}|.
\ee
The arguments based on holomorphy do not apply and therefore 
$u$-squark cannot be decoupled from the rest of the squarks.
This is sufficient to generate possibly small but non-vanishing
$m_u$ at the SUSY threshold which
turns out to be
\be
m_u \simeq \eta~ m_t~\fr{\alpha_s}{18\pi}~\fr{A-\mu\cot\beta}{m}~
\lambda^{|X_{Q_1}-X_{Q_3}|+|X_{U_1}-X_{U_3}|}.
\ee
Here again we take advantage of the possibility to ``import''  
large $m_t$ through the flavour-changing along the squark line.
$\eta\sim 2.5$ accounts for the QCD renormalization change of $m_u$
from SUSY threshold to 1 GeV. Assuming that $\lambda$ is equal to
the Wolfenstein's parameter $\lambda_W=0.22$, we arrive at the 
following bound on the combination of the horizontal charges
in the Up quark sector:
\be
|X_{Q_1}-X_{Q_3}|+|X_{U_1}-X_{U_3}| > 17.
\label{charges}
\ee
Is such a hierarchy of horizontal charges natural? 
At the very least it is another serious model-building problem. 

\section{Conclusion}

The vanishing of the theta term at the tree level might be
achieved via an exact left-right symmetry, acting above
certain high energy scale $\Lambda$. The particular form of the 
gauge group which permits such a symmetry is not 
important for the solution of the strong CP problem. 

What is crucial, however, is the degree of the universality 
in the soft-breaking sector which influences the value of 
the radiatively induced $\bar\theta$ term. Assuming  
the MSSM field structure below a certain scale $\Lambda$ and exact 
left-right symmetry above this scale, we studied  
the allowed departure from the flavour universality in the squark mass 
sector. We find that the 0.5\% mass splitting among squarks at the scale 
$\Lambda$ can be consistent with the bounds on 
$\bar\theta$, but only in two very specific cases.
The first case corresponds to a situation when the squark mass matrix can be 
diagonalized in the same basis as the Down-quark Yukawa matrix. 
The second one is when the departure from the universality is 
proportional to the combination of Up and Down Yukawa matrices. The
latter 
form of the mass matrix for squarks may result from the renormalization
group evolution of the initially universal squark masses between 
Plank scale and $\Lambda$. Thus any model-building effort which
tries to explain the smallness of $\bar \theta$ by a restoration of
parity has to ensure that squark masses fall into one of these two
patterns. 

We have also shown that a possibility of 
$m_u=0$ type of solution to the strong CP problem 
in the context of horizontal symmetries depends upon the size of 
$M_u$ generated radiatively at SUSY threshold. 
The squark-gluino exchange diagram would typically 
induce a nonzero value for 
$m_u$ which can be expressed in terms of the differences between 
horizontal charges of the quark superfields from the first and 
third generation. Assuming that the order parameter governing 
the hierarchical structure is of the order of Wolfenstein's $\lambda$,
we obtain the constraint $|X_{Q_1}-X_{Q_3}|+|X_{U_1}-X_{U_3}| > 17$ 
which represents a serious model-building challenge.

\section{Acknowledgements}
This work is supported in part by N.S.E.R.C. of Canada and DOE 
grant DE-FG02-94ER-40823 at the University of Minnesota. 
M.P. would like to thank Alon Faraggi, Rabi Mohapatra 
and Jogesh Pati for interesting stimulating discussions.

\end{document}